\documentclass{article}
\usepackage{waspaa23,amsmath, amssymb,graphicx,url,times}
\usepackage{color}
\usepackage{tikz}
\usepackage{hyperref}

\newcommand\norm[1]{\lVert#1\rVert}

\newcommand\babs[1]{\big|#1\big|}
\newcommand\floor[1]{\lfloor#1\rfloor}

\DeclareMathOperator{\relu}{ReLU}

\usepackage{tikz}
\usetikzlibrary{shapes.geometric, arrows, backgrounds}

\tikzstyle{inout} = [rectangle, rounded corners, minimum width=2cm, minimum height=0.7cm, text centered, draw=black, fill=black!10]
\tikzstyle{conv1d} = [rectangle, rounded corners, minimum width=2.5cm, minimum height=0.7cm, text centered, draw=black, fill=orange!30]
\tikzstyle{pool} = [rectangle, rounded corners, minimum width=2.5cm, minimum height=0.7cm, text centered, draw=black, fill=blue!30]
\tikzstyle{tconv1d} = [rectangle, rounded corners, minimum width=2.5cm, minimum height=0.7cm, text centered, draw=black, fill=orange!70]
\tikzstyle{gru} = [rectangle, rounded corners, minimum width=2.5cm, minimum height=0.7cm, text centered, draw=black, fill=red!70]
\tikzstyle{adaconv1d} = [rectangle, rounded corners, minimum width=2.5cm, minimum height=0.7cm, text centered, draw=black, fill=red!30!orange!30]
\tikzstyle{adacomb1d} = [rectangle, rounded corners, minimum width=2.5cm, minimum height=0.7cm, text centered, draw=black, fill=red!30!orange!60]
\tikzstyle{arrow}=[draw, -latex]

\title{LACE: A light-weight, causal model for enhancing coded speech through adaptive convolutions}

\name{Jan B\"uthe,
      Jean-Marc Valin,
      Ahmed Mustafa
      }
\address{Amazon Web Services\\ Palo Alto, USA \\
{\small \{jbuethe, jmvalin, ahdmust\}@amazon.com}
}

\begin{document}

\ninept
\maketitle

\begin{sloppy}

\begin{abstract}

Classical speech coding uses low-complexity postfilters with zero lookahead to enhance the quality of coded speech, but their effectiveness is limited by their simplicity. Deep Neural Networks (DNNs) can be much more effective, but require high complexity and model size, or added delay. We propose a DNN model that generates classical filter kernels on a per-frame basis with a model of just 300~K parameters and 100~MFLOPS complexity, which is a practical complexity for desktop or mobile device CPUs. The lack of added delay allows it to be integrated into the Opus codec, and we demonstrate that it enables effective wideband encoding for bitrates down to 6~kb/s.

\end{abstract}

\section{Introduction}
Degradation of speech through coding is a persisting problem, especially in communication scenarios where delay and complexity are critical. Classical approaches to coded speech enhancement \cite{chen_adaptive_postfiltering} are typically very low in complexity and do not require lookahead but are also of limited effectiveness due to the simplistic nature of the algorithms. A number of DNN based methods have recently been proposed. \cite{zhao_enhancement, opus_resynthesis, gupta_speech_enhancement, korse_speech_enhancement, korse_postgan}. These are more effective but they are also either high in complexity or they require additional lookahead. The reason behind this is the more general observation that time-domain models for signal enhancement, which can operate in a causal manner, tend to be much larger and more complex than frequency domain methods, which in turn require lookahead for overlap-add \cite{ochieng_dnn}. This issue becomes even more pressing when the speech codec that is to be enhanced is embedded into a larger coding structure, which is often the case for modern codecs like Opus or EVS.

As a solution to this problem we propose the linear-adaptive coding enhancer (LACE), a light-weight causal model with only 300~K parameters and 100~MFLOPS complexity. The key idea behind LACE is to use adaptive convolutions with filter kernels computed from input features on a per-frame basis at inference time. This eliminates the need to have a large number of channels, which is typical for CNNs. We also base the architecture on classic post-filters \cite{chen_adaptive_postfiltering}, providing the model with a large but sparse receptive field where needed, which reduces complexity even further.

We test our model by applying it to the linear-predictive coding mode of the Opus codec, also known as SILK \cite{koen_silk, rfc6716}\footnote{samples are available at \href{https://282fd5fa7.github.io/LACE}{https://282fd5fa7.github.io/LACE}}. Since LACE does not require lookahead and is essentially phase preserving, it can be directly integrated into the decoder while maintaining the seemless mode-switching capability. The model is bitrate-scalable, and we verify in a P.808 listening test that it significantly improves the baseline at 6, 9 and 12 kb/s. We also provide PESQ scores up to 22 kb/s, which demonstrate that LACE scales to transparency as the codec does. The low complexity and size also allow the model to run on both desktop and mobile device CPUs with insignificant overhead making it a practical and effective method.

Although we chose Opus as a test case for LACE, the model could be easily adapted to any speech codec that provides pitch information at the decoder side. Compared to fully neural codecs \cite{valin_lpcnet_coding, kleijn_lyra, zegidhour_soundstream,pia_nesc, jenrungrot_lmcodec}, this approach has the practical advantage of maintaining backward compatibility, leaving an inexpensive decoding option for low-end devices like microcontrollers.

\section{LACE}
The task of enhancing coded speech is loosely related to a denoising problem, i.e. recovering a clean signal $x(t)$ from a noisy mixture
\begin{equation}\label{e:noisy-mixture}
y(t) = x(t) + n(t).
\end{equation}
However, for a speech codec (or any other perceptual codec) the coding noise $n(t)$ will be closely related to the signal
$x(t)$ itself to exploit masking properties. This means that recovering $x(t)$ from $y(t)$ is neither a feasible nor desirable task. In particular,
training a model to minimize the mean-square error between enhanced signal and clean signal will result in losing large parts of the speech signal
since noise in $x(t)$ is largely replaced by statistically similar noise in $y(t)$. We therefore state the task as a noise-reshaping task. Given the noisy mixture in \eqref{e:noisy-mixture} we want to produce an enhanced signal $\hat y(t)$ in which the coding noise is less audible.

Classical approaches \cite{chen_adaptive_postfiltering} identify spectral valleys as main source for audible coding noise. These include both
the narrow valleys between harmonics for voiced speech parts as well as wider valleys between formants, the peaks of the spectral envelope.
The first task, inter-harmonic noise reduction, is classically addressed by a long term post filter, a comb-filter that makes explicit use of the pitch lag, emphasizing multiples of the fundamental frequency and attenuating frequencies in between. The second task, formant enhancement, is carried out with a short-term-filter, usually derived from the short-term linear prediction coefficients.

Combining these ideas naturally results in time-varying filtering model
\begin{equation}\label{d:yhat-iir}
\hat y(t) = \sum_{\tau = 0}^{\infty} h(t, \tau) \, y(t - \tau),
\end{equation}
which we take as starting point for our investigations. By computing these filters from input data we can not only apply formant and pitch enhancement to the coded signal but the model can also address temporal artifacts, which is clearly beyond the reach of classical post-filters, which are capable of only limited adaptation. The approach differs significantly from standard non-adaptive CNNs like the time-domain model in \cite{zhao_enhancement}, which compensate for that lack of adaptation by using a large number of redundant channels, making the models both large and computationally expensive.
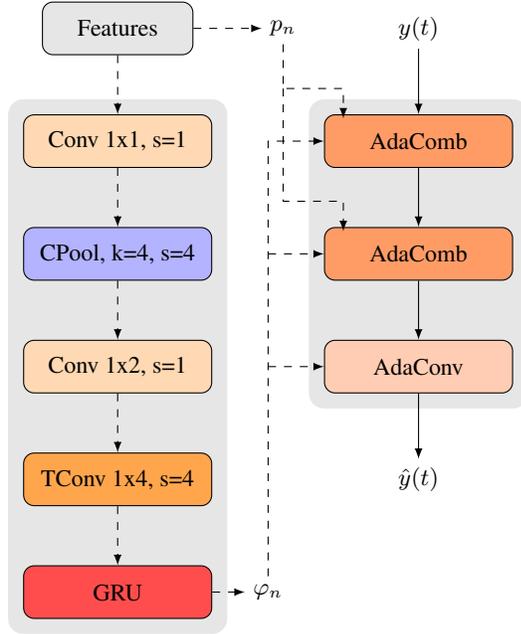
\begin{figure}
\center
\scalebox{1}{
\begin{tikzpicture}[node distance=1.2cm]

\node (features) [inout] at (0, 0) {Features};
\node (fconv1) [conv1d] at (0, -1.5) {Conv 1x1, s=1};
\node (fpool) [pool] at (0, -3) {CPool, k=4, s=4};
\node (fconv2) [conv1d] at (0, -4.5) {Conv 1x2, s=1};
\node (ftconv) [tconv1d] at (0, -6) {TConv 1x4, s=4};
\node (fgru) [gru] at (0, -7.5) {GRU};
\node (phi) at (2, -7.5) {\small$\varphi_n$};

\node (lags) at (2.2, 0) {$p_n$};

\node (sigin)  at (4, 0) {$y(t)$};
\node (adacomb1) [adacomb1d] at (4, -1.5) {AdaComb};
\node (adacomb2) [adacomb1d] at (4, -3) {AdaComb};
\node (adaconv) [adaconv1d] at (4, -4.5) {AdaConv};
\node (denoised)  at (4, -6) {$\hat y(t)$};

\draw [arrow, dashed] (features) -- (fconv1);
\draw [arrow, dashed] (fconv1) -- (fpool);
\draw [arrow, dashed] (fpool) -- (fconv2);
\draw [arrow, dashed] (fconv2) -- (ftconv);
\draw [arrow, dashed] (ftconv) -- (fgru);

\draw [arrow, dashed] (fgru) -- (phi);
\draw [arrow, dashed] (phi) -- (2, -1.5) -- (adacomb1);
\draw [arrow, dashed] (2, -3) -- (adacomb2);
\draw [arrow, dashed] (2, -4.5) -- (adaconv);

\draw [arrow] (sigin) -- (adacomb1);
\draw [arrow] (adacomb1) -- (adacomb2);
\draw [arrow] (adacomb2) -- (adaconv);
\draw [arrow] (adaconv) -- (denoised);

\draw [arrow, dashed] (features) -- (lags);
\draw [arrow, dashed] (lags) -- (2.2, -0.8) -- (3, -0.8) -- (3, -1.2);
\draw [arrow, dashed] (2.2, -0.8) -- (2.2, -2.3) -- (3, -2.3) -- (3, -2.7);

\begin{pgfonlayer}{background}
\filldraw [line width=4mm, join=round, black!10]
(fconv1.north -| fconv1.east) rectangle (fgru.south -| fgru.west)
(adacomb1.north -| adacomb1.east) rectangle (adaconv.south -| adaconv.west);
\end{pgfonlayer}{background}

\end{tikzpicture}
}
\caption{High-level overview of LACE model. The feature encoder on the left produces latent feature vectors $\varphi_n$ every 5~ms. These feature vectors are used by the adaptive filtering modules in the signal path on the right to calculate convolution kernels. In addition, the comb-filtering modules make direct use of the pitch lags $p_n$.}\label{f:blockdiagram}
\end{figure}
\subsection{Model Overview}
The LACE model implements a finite version of \eqref{d:yhat-iir} with a varying filter length at most roughly twice the maximal pitch lag, which is depicted as signal path in Fig. \ref{f:blockdiagram}. Computing the filter coefficients directly would be possible but inefficient. Instead we apply multiple consecutive (comb-)filtering modules which result in sparse filters for large pitch lags. The filtering modules are described in detail in section \ref{s:adaptive-convolution} .

The filtering modules in the signal path are steered with features derived from the Opus decoder which are transformed into a sequence of latent feature vectors $\varphi_n$ by a feature encoder depicted in the left of Fig. \ref{f:blockdiagram}. These latent feature vectors are calculated at a rate of 200~Hz, which corresponds to the sub-frame rate of the Opus linear-predictive coding mode. The feature processing is causal in the sense that it never uses information beyond the current 20-ms Opus frame.

\subsection{Features}
To keep complexity low and model size small, we base our model on a set of carefully hand-tuned features. As in \cite{opus_resynthesis}, these are a mix of (quantized) clean speech features received by the Opus decoder and features calculated from the noisy decoded speech.

For clean speech features we use
\begin{enumerate}
\item LPC coefficients converted to a 64-band ERB-scale log-magnitude spectrogram at 10-ms granularity
\item Opus quantization gains in log-domain at 5-ms granularity
\item pitch lags at 5-ms granularity
\item 5 filter taps from the Opus LTP filter at 5-ms granularity
\end{enumerate}
and for noisy speech features we use
\begin{enumerate}
\item 18-band cepstrum on 20-ms frames at a granularity of 10~ms following the Opus CELT bands
\item 5 auto-correlation values of $y(t)$ around the Opus pitch lag at a granularity of 5~ms
\end{enumerate}

The features are selected to be largely scale-invariant, the two exceptions being the Opus frame gain and the constant term in the noisy speech cepstrum. The reason for this is that the filtering in \eqref{d:yhat-iir} as a linear operation should not depend on the signal level. Furthermore, for simplicity all features are upsampled to 200~Hz by repetition.

\subsection{Bitrate Signalling}
Since the kind and strength of coding artifacts strongly depends on the bitrate, it is essential to provide this information to the LACE model. However, since the linear-predictive Opus mode is a variable bitrate encoder by design, the exact target bitrate specified at the encoder is not known to the decoder. We let the model infer the bitrate from the number of bits of the received Opus frames. Since these values give a very noisy estimate of the bitrate, we provide both the raw number of bits and an exponential moving average with an update rate on $0.1$  to the feature encoder.

To make these features more useful, we pass them through an eight-dimensional saturating embedding whose components are given by
\begin{equation}
\sin\left( k \, \frac{2 \log(\max\{A,\, \min\{B, n_{bits}\}\}) - \log(A\cdot B)}{\log(B/A)}\right)
\end{equation}
for $k=1, 2, \dots, 8$.
The reasoning behind the saturation is that there is nothing to do for the model at very high bitrates. Hence, it suffices to add a single such high bitrate to the training data to ensure that the model behaves well outside of the intended range of use. We use $A=50$ and $B=650$, which correspond to bitrates of $2.5$ and $32.5$ kb/s.

\subsection{Feature Encoding}
Feature encoding is done by a sequential model of convolutions and transpose convolutions with $\tanh$ activations followed by a GRU to capture long-term dependencies. The first convolution performs scaling and dimensionality reduction on the 178-dimensional feature space (a 64-dimensional embedding is used for the pitch values), reducing the features to $N_r=96$ channels. This is followed by a concatenative pooling layer, which combines the information from the four Opus subframes into one feature vector. This is followed by a second convolution, which adjusts the number of channels to the hidden feature dimension $N_h$ and a transpose convolution with $N_h$ channels that performs a factor 4 upsampling back to 200~Hz. Finally, the result is processed by a GRU with $N_h$ hidden units to generate the $N_h$-dimensional hidden feature vectors $\varphi_n$. The pooling and upsampling provides the model with partial lookahead to the Opus frame boundary. The model is therefore causal on the Opus frame level but not causal on the Opus sub-frame level.

\subsection{Adaptive Convolutions}\label{s:adaptive-convolution}
The AdaptiveConv module in Fig. \ref{f:blockdiagram} filters the input signal with a sequence of FIR filters. It plays the role of formant enhancement in the LACE model.

The FIR filter coefficients are calculated on a per-frame basis from the hidden feature vectors $\varphi_n$ as
\begin{equation}
h_n(\tau) = g_n \, \kappa_n(\tau),
\end{equation}
where $\kappa_n$ is the filter shape calculated as
\begin{equation}
\kappa_n = \frac{W_\kappa \varphi_n + b_\kappa}{\norm{W_\kappa \varphi_n + b_\kappa}_2}.
\end{equation}
and $g_n$ is the filter gain calculated as
\begin{equation}
g_n = \exp(\alpha \tanh(W_g \varphi_n + b_g))
\end{equation}
with trainable projection matrices $W_\cdot$ and biases $b_\cdot$. Splitting the filter coefficients into shape and gain is similar to weight normalization \cite{salimans_weightnorm} for regular convolutional layers but we also use it to limit the maximal amplification of the filters.

To avoid transition artifacts, the filters $h_n$ are interpolated on the first half of the 5-ms frames using a half Hann window.

The AdaptiveComb modules in Fig. \ref{f:blockdiagram} are similar to the AdaptiveConv but the FIR filter taps are moved around the pitch delay $p_n$. Furthermore, they feature a second gain to control the comb-filtering strength, which is calculated as
\begin{equation}
 \gamma_n = \exp(\beta - \relu (W_\gamma \varphi_n + b_\gamma)).
\end{equation}
We use a ReLU activation to calculate $\gamma_n$ since we only need a one-sided limitation on the log-scale to limit the comb-filtering strength. The transfer function on frame $n$ is thus given by
\begin{equation}\label{e:comb-filter-transfer}
H_n(z) = g_n\left( 1 + \gamma_n z^{-p_n + \floor{k / 2}} \sum_{\ell = 0 }^{k-1} \kappa_n(\ell) z^{-\ell} \right),
\end{equation}
where $k$ denotes the filter length. The filters are again interpolated with a half Hann window.

Comb filtering is only useful for voiced speech parts and Opus does not transmit a pitch lag for speech parts classified as unvoiced by the encoder. For such frames we set the pitch lag to $p_n = \floor{k / 2}$, essentially turning the AdaptiveComb modules into AdaptiveConv modules. This limits the risk of adding coloration to unvoiced speech parts and increases the spectral shaping capability.

\section{Training}
\subsection{Data}
We train our model on 165 hours of clean speech sampled at 16~kHz collected from multiple high-quality TTS datasets \cite{demirsahin-etal-2020-open, kjartansson-etal-2020-open, kjartansson-etal-tts-sltu2018, guevara-rukoz-etal-2020-crowdsourcing, he-etal-2020-open, oo-etal-2020-burmese, van-niekerk-etal-2017, gutkin-et-al-yoruba2020, bakhturina21_interspeech}, which contains more than 900 speakers in 34 languages and dialects. The data is augmented using random scaling and random equalization. We obtain the coded signal with a modified version of libopus, restricting the encoder to wideband and linear-predictive mode only. Furthermore, we change the encoder parameters complexity, packet\_loss\_percent and bitrate randomly every 249 frames. Since Opus applies a high-pass filter to the clean signal before encoding in linear-predictive mode, we use that high-pass filtered clean signal as the target. Furthermore, we apply pre-emphasis with $P(z) = 1 - 0.85 z^{-1}$ to both the noisy input signal and the clean target signal. We choose hyper parameters $N_r=96$, $N_h=128$ and $k=15$ which results in a model size of $306$~K parameters and $99$~MFLOPS complexity.

\subsection{Loss}
We use a mixture of regression losses to train the LACE model, which are tailored to different tasks. One loss is calculated in the time domain and the other two losses are calculated on STFTs with window size equal to DFT size and $50\%$ overlap. The STFT losses are averaged over different resolutions with DFT sizes $2^n$, $n=5, 6, \dots, 12$. All STFTs are calculated with Hann windows.

The LACE model is not intrinsically phase preserving. As a matter of fact the only zero-phase filter it can implement is the identity. We therefore bias towards phase preservation in periodic signal parts by applying a weighted squared error loss
\begin{equation}\label{e:weighted-mse}
\mathcal{L}_{\mathrm{phase}} = \frac{\norm{x - \hat y}_2^2}{\norm{\hat y}_2}.
\end{equation}
The weighting by $1/\norm{\hat y}_2$ modifies the $L^2$ loss behavior on unvoiced signal parts. We pointed out before, that the MSE loss will lead to loss of unvoiced signal parts since $x(t)$ and $y(t)$ will be largely uncorrelated. However, for uncorrelated $x(t)$ and $y(t)$ the weighted version \eqref{e:weighted-mse} attains its global minimum at $\norm{x}_2 = \norm{\hat y}_2$ (as opposed to $\norm{\hat y} = 0$ for the unweighted MSE loss) which makes the loss energy preserving in this case.

For envelope reconstruction, we apply a set of perceptually motivated filters to smooth the absolute values of the STFT coefficients. The filters are approximations to the auditory filters as described in \cite{moore2012introduction} following an ERB scale. As envelope loss we use the L1 loss on the resulting smooth spectrograms $X_s$ and $\hat Y_s$, i.e.
\begin{equation}
\mathcal{L}_{\mathrm{env}}= \norm{\log(X_s) - \log(\hat {Y_s})}_1.
\end{equation}

To restore the harmonic structure, we use a modification of the spectral convergence loss, where we replace the Frobenius norm by a cross-correlation based loss, which makes it insensitive to signal scale. The cross-correlation is calculated both over time and frequency, giving
\begin{equation}
\mathcal{L}_{\mathrm{spec}} = 1 -  \frac{\sum_{n, k} \babs{X(n, k)} \babs{\hat Y(n, k)}}{\left(\sum_{n, k} \babs{ X(n, k)}^2  \sum_{n, k} \babs{\hat Y(n, k)}^{2} \right)^{1/2}}.
\end{equation}

As total loss we use
\begin{equation}\label{e:total-loss}
\mathcal{L}_{\mathrm{total}} = 10 \, \mathcal{L}_{\mathrm{phase}}  + 2\, \mathcal{L}_{\mathrm{env}} + \mathcal{L}_{\mathrm{spec}}
\end{equation}

\subsection{Training Details}
We train the model using the Adam optimizer with $\beta_1=0.9$ and $\beta_2=0.999$ on sequences of $0.5$~s length with a mini-batch size of $256$ and an initial learning rate $\lambda = 5\times 10^{-4}$ for 50 epochs or $\approx 230$ steps. We use learning rate decay with a factor $2.5\times 10^{-5}$, i.e. in step $\mu$ the weights are updated with a learning rate $\lambda / (1 + 2.5\times 10^{-5}\, \mu)$.

\section{Evaluation}
\subsection{Frequency Response Analysis}
Since the LACE signal path is linear we can study local model behavior by analysing frequency responses. Fig. \ref{f:freqz} shows a series of frequency responses for a voiced frame. For low bitrates LACE performs strong comb filtering and for higher bitrates it approaches the identity function, which is the intended behavior.
\begin{figure}
\centering
\includegraphics[scale=1]{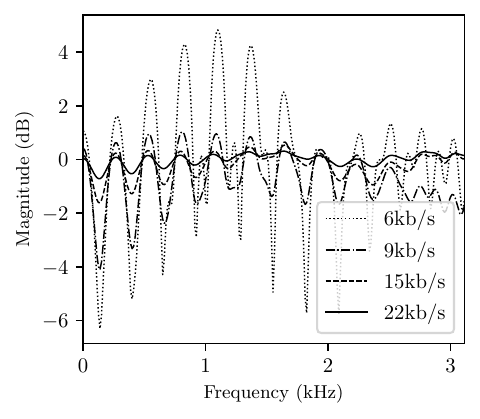}
\caption{Frequency responses of the LACE signal path for a voiced frame at multiple bitrates}\label{f:freqz}
\end{figure}

\subsection{Listening Test}

We evaluated the quality of LACE using 192~clean English speech clips from the NTT Multi-Lingual Speech Database for Telephonometry, which was not included in the training data. Using the crowd-sourcing methodolgy from ITU-R P.808~[19], we tested Opus with and without LACE at 6, 9, and 12~kb/s. 

As a benchmark for causal coded speech enhancement, we included the TDCNN time domain method from~[2]. Since the model is not designed to operate with multiple bitrates, we restrict training and comparison to 6~kb/s. We adapted the TDCNN training setup and hyper-parameters, increasing its PESQ score by 0.2, and lowering its complexity. We used $N=15$, $F=55$, and $L=320$, and improved the loss function by using~(12) instead of MSE.

Although it cannot be used in a real-time Opus implementation due to its 25~ms delay, we also included the non-causal LPCNet resynthesis method from~[3] as a comparison point. Such resynthesis methods are also limited to enhancing lower bitrates (6~kb/s in this case), since their output quality is bounded by the quality of the vocoder, which prevents them from scaling to transparency. The exact bitrate threshold, however, will depend on the vocoder.

Results in Fig.~3 show that LACE significantly improves Opus at all tested bitrates. At 6~kb/s LACE outperforms TDCNN by a large margin and achieves about 60\% of the quality improvement of the non-causal resynthesis method. LACE has a total complexity of 100~MFLOPS (0.1~GFLOPS), far lower than the 3~GFLOPS required for resynthesis and the 16~GFLOPS of TDCNN.

\begin{figure}
\centering
\includegraphics[scale=1]{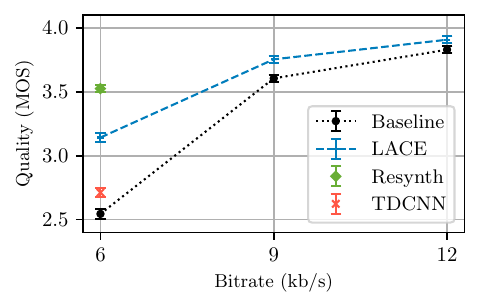}
\caption{P.808 results. The clean signal has a MOS of $4.01\pm 0.03$. LACE consistently outperforms the baseline and TDCNN and achieves about $60\%$ of the MOS improvement of LPCNet resynthesis at 6kb/s which requires 25ms delay at 4x the size and 30x the complexity compared to LACE.}
\end{figure}

\subsection{Objective Evaluation}
We use PESQ to test performance of LACE from 6 to 22 kb/s. As can be seen in Fig. \ref{fig:pesq} LACE always outperforms the baseline (Opus). Quantitatively, the PESQ improvement is in line with the listening test results at 6, 9 and 12 kb/s.
\begin{figure}
\centering
\includegraphics[scale=1]{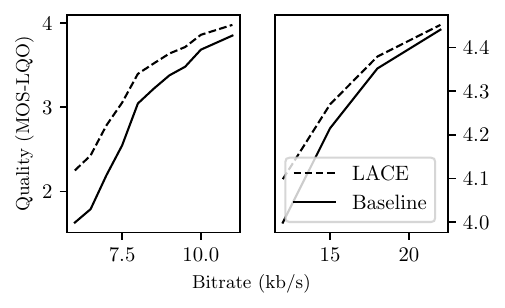}
\caption{PESQ scores from 6 to 22~kb/s.}\label{fig:pesq}
\end{figure}

\section{Conclusion}
We have demonstrated that adaptive convolutions give rise to a very efficient and effective enhancer for coded speech. LACE consistently improves the test codec over a large bitrate range, demonstrated both by objective and subjective methods, and it outperforms state-of-the-art causal methods by a large margin. Since LACE is both causal and essentially phase preserving, it can be directly integrated into codecs with dedicated speech-coding modes. We believe such a low-complexity enhancement algorithm will be most useful for enhancing the quality of existing classical speech codecs without breaking compatibility.

\section{ACKNOWLEDGMENT}
\label{sec:ack}

The authors would like to thank Timothy Terriberry, Paris Smaragdis and Mike Goodwin for helpful suggestions.

% -------------------------------------------------------------------------
% Either list references using the bibliography style file IEEEtran.bst
\bibliographystyle{IEEEtran}
\bibliography{audio}

\begin{thebibliography}{10}
\providecommand{\url}[1]{#1}
\csname url@samestyle\endcsname
\providecommand{\newblock}{\relax}
\providecommand{\bibinfo}[2]{#2}
\providecommand{\BIBentrySTDinterwordspacing}{\spaceskip=0pt\relax}
\providecommand{\BIBentryALTinterwordstretchfactor}{4}
\providecommand{\BIBentryALTinterwordspacing}{\spaceskip=\fontdimen2\font plus
\BIBentryALTinterwordstretchfactor\fontdimen3\font minus
  \fontdimen4\font\relax}
\providecommand{\BIBforeignlanguage}[2]{{%
\expandafter\ifx\csname l@#1\endcsname\relax
\typeout{** WARNING: IEEEtran.bst: No hyphenation pattern has been}%
\typeout{** loaded for the language `#1'. Using the pattern for}%
\typeout{** the default language instead.}%
\else
\language=\csname l@#1\endcsname
\fi
#2}}
\providecommand{\BIBdecl}{\relax}
\BIBdecl

\bibitem{chen_adaptive_postfiltering}
J.-H. Chen and A.~Gersho, ``{Adaptive Postfiltering for Quality Enhancement of
  Coded Speech},'' \emph{IEEE Transactions on Speech and Audio Processing},
  vol.~3, no.~1, pp. 59--71, 1995.

\bibitem{zhao_enhancement}
Z.~Zhao, H.~Liu, and T.~Fingscheidt, ``{Convolutional Neural Networks to
  Enhance Coded Speech},'' \emph{IEEE/ACM Transactions on Audio, Speech, and
  Language Processing}, vol.~27, no.~4, pp. 663--678, 2019.

\bibitem{opus_resynthesis}
J.~Skoglund and J.-M. Valin, ``{Improving Opus Low Bit Rate Quality with Neural
  Speech Synthesis},'' in \emph{Proc. INTERSPEECH}, 2019.

\bibitem{gupta_speech_enhancement}
K.~Gupta, S.~Korse, B.~Edler, and G.~Fuchs, ``{A DNN Based Post-Filter to
  Enhance the Quality of Coded Speech in MDCT Domain},'' in \emph{Proc.
  International Conference on Acoustics, Speech and Signal Processing
  (ICASSP)}, 2022, pp. 836--840.

\bibitem{korse_speech_enhancement}
S.~Korse, K.~Gupta, and G.~Fuchs, ``{Enhancement of Coded Speech Using a
  Mask-Based Post-Filter},'' in \emph{Proc. International Conference on
  Acoustics, Speech and Signal Processing (ICASSP)}, 2020, pp. 6764--6768.

\bibitem{korse_postgan}
S.~Korse, N.~Pia, K.~Gupta, and G.~Fuchs, ``{PostGAN: A GAN-Based
  Post-Processor to Enhance the Quality of Coded Speech},'' in \emph{Proc.
  International Conference on Acoustics, Speech and Signal Processing
  (ICASSP)}, 2022, pp. 831--835.

\bibitem{ochieng_dnn}
P.~Ochieng, ``{Deep Neural Network Techniques for Monaural Speech Enhancement:
  State of the Art Analysis},'' \emph{arXiv:2212.00369}, 2022.

\bibitem{koen_silk}
K.~Vos, K.~Sørensen, S.~Jensen, and J.-M. Valin, ``{Voice Coding with Opus},''
  \emph{135th AES Convention}, pp. 722--731, 2013.

\bibitem{rfc6716}
J.-M. Valin, K.~Vos, and T.~B. Terriberry, ``{Definition of the Opus Audio
  Codec},'' RFC 6716, 2012.

\bibitem{valin_lpcnet_coding}
J.-M. Valin and J.~Skoglund, ``{A Real-Time Wideband Neural Vocoder at 1.6kb/s
  Using LPCNet},'' in \emph{Proc. INTERSPEECH}, 2019, pp. 3406--3410.

\bibitem{kleijn_lyra}
W.~B. Kleijn, A.~Storus, M.~Chinen, T.~Denton, F.~S.~C. Lim, A.~Luebs,
  J.~Skoglund, and H.~Yeh, ``{Generative Speech Coding with Predictive Variance
  Regularization},'' in \emph{Proc. International Conference on Acoustics,
  Speech and Signal Processing (ICASSP)}, 2021, pp. 6478--6482.

\bibitem{zegidhour_soundstream}
N.~Zeghidour, A.~Luebs, A.~Omran, J.~Skoglund, and M.~Tagliasacchi,
  ``{SoundStream: An End-to-End Neural Audio Codec},'' \emph{IEEE/ACM
  Transactions on Audio, Speech, and Language Processing}, vol.~30, pp.
  495--507, 2022.

\bibitem{pia_nesc}
N.~Pia, K.~Gupta, S.~Korse, M.~Multrus, and G.~Fuchs, ``{NESC: Robust Neural
  End-2-End Speech Coding with GANs},'' in \emph{Proc. INTERSPEECH}, 2022.

\bibitem{jenrungrot_lmcodec}
T.~Jenrungrot, M.~Chinen, W.~Kleijn, J.~Skoglund, Z.~Borsos, N.~Zeghidour, and
  M.~Tagliasacchi, ``Lmcodec: A low bitrate speech codec with causal
  transformer models,'' in \emph{Proc. International Conference on Acoustics,
  Speech and Signal Processing (ICASSP)}, 2023.

\bibitem{salimans_weightnorm}
T.~Salimans and D.~P. Kingma, ``{Weight Normalization: A Simple
  Reparameterization to Accelerate Training of Deep Neural Networks},'' in
  \emph{NIPS}, 2016.

\bibitem{demirsahin-etal-2020-open}
I.~Demirsahin, O.~Kjartansson, A.~Gutkin, and C.~Rivera, ``{Open-source
  Multi-speaker Corpora of the English Accents in the British Isles},'' in
  \emph{Proc. LREC}, 2020.

\bibitem{kjartansson-etal-2020-open}
O.~Kjartansson, A.~Gutkin, A.~Butryna, I.~Demirsahin, and C.~Rivera,
  ``{Open-Source High Quality Speech Datasets for Basque, Catalan and
  Galician},'' in \emph{Proc. SLTU and CCURL}, 2020.

\bibitem{kjartansson-etal-tts-sltu2018}
K.~Sodimana, K.~Pipatsrisawat, L.~Ha, M.~Jansche, O.~Kjartansson, P.~D. Silva,
  and S.~Sarin, ``{A Step-by-Step Process for Building TTS Voices Using Open
  Source Data and Framework for Bangla, Javanese, Khmer, Nepali, Sinhala, and
  Sundanese},'' in \emph{Proc. SLTU}, 2018.

\bibitem{guevara-rukoz-etal-2020-crowdsourcing}
A.~Guevara-Rukoz, I.~Demirsahin, F.~He, S.-H.~C. Chu, S.~Sarin,
  K.~Pipatsrisawat, A.~Gutkin, A.~Butryna, and O.~Kjartansson, ``{Crowdsourcing
  Latin American Spanish for Low-Resource Text-to-Speech},'' in \emph{Proc.
  LREC}, 2020.

\bibitem{he-etal-2020-open}
F.~He, S.-H.~C. Chu, O.~Kjartansson, C.~Rivera, A.~Katanova, A.~Gutkin,
  I.~Demirsahin, C.~Johny, M.~Jansche, S.~Sarin, and K.~Pipatsrisawat,
  ``{Open-source Multi-speaker Speech Corpora for Building Gujarati, Kannada,
  Malayalam, Marathi, Tamil and Telugu Speech Synthesis Systems},'' in
  \emph{Proc. LREC}, 2020.

\bibitem{oo-etal-2020-burmese}
Y.~M. Oo, T.~Wattanavekin, C.~Li, P.~De~Silva, S.~Sarin, K.~Pipatsrisawat,
  M.~Jansche, O.~Kjartansson, and A.~Gutkin, ``{Burmese Speech Corpus,
  Finite-State Text Normalization and Pronunciation Grammars with an
  Application to Text-to-Speech},'' in \emph{Proc. LREC}, 2020.

\bibitem{van-niekerk-etal-2017}
D.~van Niekerk, C.~van Heerden, M.~Davel, N.~Kleynhans, O.~Kjartansson,
  M.~Jansche, and L.~Ha, ``{Rapid development of TTS corpora for four South
  African languages},'' in \emph{Proc. INTERSPEECH}, 2017.

\bibitem{gutkin-et-al-yoruba2020}
A.~Gutkin, I.~Demir{\c{s}}ahin, O.~Kjartansson, C.~Rivera, and
  K.~T{\'u}b\d{\`o}s{\'u}n, ``{Developing an Open-Source Corpus of Yoruba
  Speech},'' in \emph{Proc. INTERSPEECH}, 2020.

\bibitem{bakhturina21_interspeech}
E.~Bakhturina, V.~Lavrukhin, B.~Ginsburg, and Y.~Zhang, ``{Hi-Fi Multi-Speaker
  English TTS Dataset},'' in \emph{Proc. INTERSPEECH}, 2021, pp. 2776--2780.

\bibitem{moore2012introduction}
B.~Moore, \emph{{An Introduction to the Psychology of Hearing}}.\hskip 1em plus
  0.5em minus 0.4em\relax Brill, 2012.

\end{thebibliography}

\end{sloppy}
\end{document}